\documentclass{article}
\usepackage{spconf,amsmath,graphicx}

\title{Personalized Automatic Speech Recognition Trained on Small Disordered Speech Datasets}

\name{Jimmy Tobin and Katrin Tomanek}
\address{Google Research, USA}

\begin{document}

\maketitle

\begin{abstract}
This study investigates the performance of personalized automatic speech recognition (ASR) for recognizing disordered speech using small amounts of per-speaker adaptation data. 
We trained personalized models for 195 individuals with different types and severities of speech impairment with training sets ranging in size from $<$1 minute to 18-20 minutes of speech data. 
Word error rate (WER) thresholds were selected to determine Success Percentage (the percentage of personalized models reaching the target WER) in different application scenarios. For the home automation scenario, 79\% of speakers reached the target WER with 18-20 minutes of speech; but even with only 3-4 minutes of speech, 63\% of speakers reached the target WER.
Further evaluation found similar improvement on test sets with conversational and out-of-domain, unprompted phrases. Our results demonstrate that with only a few minutes of recordings, individuals with disordered speech could benefit from personalized ASR.
\end{abstract}
\begin{keywords}
automatic speech recognition, speech disorders, personalized models
\end{keywords}
\section{Introduction}
\label{sec:intro}

Voice controlled home automation technology offers convenience for many; however, often the automatic speech recognition (ASR) systems that power these technologies do not work for the millions of individuals with speech impairments
\cite{bhattacharyya2014prevalence}. This population could arguably benefit the most from home automation and other voice controlled assistive technology. People who have speech impairments often also have mobility impairments, due to their conditions. While ASR accuracy for typical speech may be as high as 95\% \cite{stonjic_taylor_kardas_kerkez_viaud_saravia_cucurull_2021, synnaeve_2021}, accuracy drops off significantly for disordered speech with varying levels of impairment severity and intelligibility \cite{moore18_interspeech}. 

Adaptation of speaker independent ASR systems with dysarthric speech in \cite{mustafa2014} has shown promising results, but they cite the dearth of speech data as a major hurdle. Recent work has shown the potential of personalizing ASR models for recognizing disordered speech \cite{green2021,Shor2019,Zhu2019,Gale2019,parrotron}. However, often-times these promising results are based on relatively
large amounts of speech recordings per speaker (often in the range of hours). For people with speech impairment, recording so many speech samples is often impractically difficult.

For example, ~\cite{green2021} reports average WER improvements of 75\% through model personalization on the Euphonia corpus ~\cite{macdonald2021}, a large corpus with speech recordings from people with speech impairment. While these WER improvements are very promising, it should be noted that this comes with significant recording times per speaker: The median number of utterances per speaker in that study is 1529, which on average results in about 2 hours of speech recordings. Recording speech data (e.g. triggering recording start/stops and flipping to the next prompt) is a significant investment of time and effort, especially for individuals with motor or cognitive impairments. It has been reported that 4-7 hours of recording time were required to contribute 1500 phrases ~\cite{macdonald2021}.
Two recent studies \cite{Shor2019,tomanek2021ondevice} suggest that large WER improvements may be achieved with much smaller amounts of data. However, both studies don't analyze in greater detail how much data is actually needed, how different levels of severity of speech impairment may impact the amount of data required or implications for practical applications.

This paper aims to close this gap. We analyze how much speech data from people with disordered speech is required for model personalization to achieve low enough Word Error Rates (WER) to be usable for various voice technology applications and human-in-the-loop conversations. 
Our ultimate goal is to make technology usable with minimal possible effort demanded of the speaker. Moreover, understanding how much speech data is required for model personalization can help accurately set expectations for ASR performance and better utilize the limited time and energy of people with speech impairments.

\section{Methods}
\label{sec:Methods}

\subsection{Dataset}

\begin{figure}[t]
    \centering
    \includegraphics[width=0.45\textwidth]{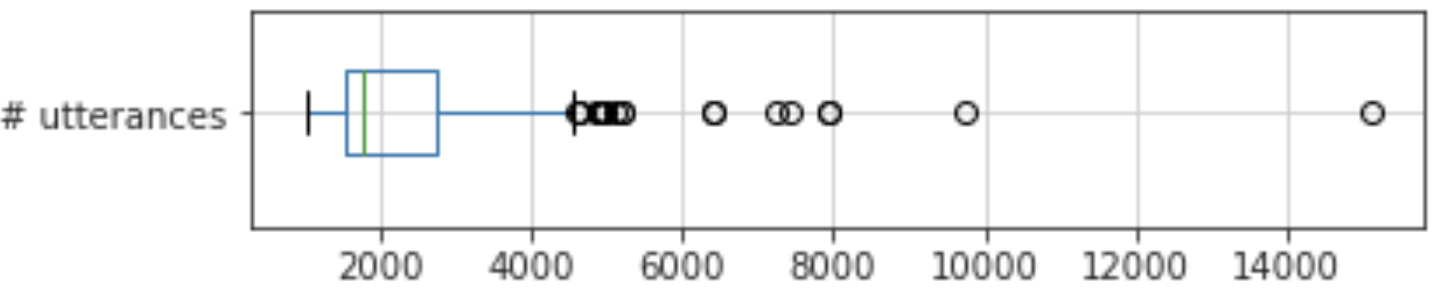}
    \caption{Total number of training utterances across all 195 speakers. Median: 1788 utterances (2.1 hours). }
    \label{fig:utt_count_boxplot}
\end{figure}

We use a subset of the Euphonia corpus~\cite{macdonald2021}. This corpus consists of over 1 million samples (over 1300 hours) of more than 1000 anonymized speakers with different types and severity levels of speech impairments. 
All our experiments are performed on a subset of 195 speakers who have each recorded more than 1000 utterances. The resulting subset is diverse in terms of severity of speech impairment (7.7\% typical, 35.4\% mild, 31.8\% moderate, 25.1\% severe) and covered etiologies (39.2\% with amyotrophic lateral sclerosis (ALS), 13.4\% cerebral palsy, 10.8\% Parkinson's Disease, 5.2\% hearing impairment, 3.1\% Ataxia etc).
We use the predefined per-speaker train, dev, and test splits (80\%/10\%/10\%) of the Euphonia corpus, which ensures that there is no phrase overlap between these splits. The median number of utterances per speaker amounts to 1788 for these 195 speakers, but variance here is high (see Figure~\ref{fig:utt_count_boxplot}).
For our study, subsets of the training data of various sizes (10, 30, 50, 100, 250 utterances) were created by randomly sampling from all utterances available per speaker. For each size, 10 subsets were created by repeated random sampling. Subsets can overlap due to sampling with replacement between sets. 
Table~\ref{tab:dataset} shows the resulting subsample sizes along with the average duration in minutes.

\begin{table}[t]
  \centering
  \begin{tabular}{|r|r|}
  \hline
  Number of  &  Average duration \\
  utterances & in minutes (std dev) \\
  \hline
  250       & 18.1 (6.6) \\
  100       & 7.3 ( 2.7) \\
  50        & 3.6 (1.3) \\
  30        & 2.2 (0.8) \\
  10        & 0.7 (0.3) \\
  \hline
    \end{tabular}
  \caption{Number of utterances and average duration of recordings averaged across speakers.}
  \label{tab:dataset}
\end{table}

\subsection{ASR Model Personalization}

For all our experiments, we employ the Recurrent Neural Network Transducers (RNN-T) architecture \cite{graves2013sequence,he2019streaming,sainath2020streaming} which allows deployment on mobile devices, supports streaming, and has demonstrated high performance.
We use 8 LSTM encoder layers and 2 LSTM layers for the language model component. Input features are 128-dimensional log Mel features computed every 10 milliseconds. 4 consecutive features are stacked with a stride of 3 frames to yield a 512-dimensional input to the encoder every 30 milliseconds. There are 4096 output units that correspond to the word piece tokens. 

We follow the same fine-tuning recipe as described in ~\cite{green2021}: We start from a speaker-independent base model pre-trained on 162k hours of typical speech. This base model has been optimized to (a) be robust across various application domains and acoustic conditions, and (b) generalize well to unseen conditions \cite{narayanan2019}. We only update the first 5 encoder layers. SpecAugment \cite{park2019} is used for data augmentation. Models were trained with a maximum number of steps based on training set size: 5000 steps for sets with $<$100 utterances, 10000 steps for 100 utterances, and 15000 steps for 250 utterances. The best checkpoint was picked using each speaker's dev set.

\subsection{Evaluation}
\label{ss:eval}
For all our experiments, we use Euphonia's default test set splits, which for the selected speakers consist of at least 100 utterances. 
We report WER on two specific domains of the Euphonia corpus: phrases associated with a) home automation\footnote{ These are short phrases of 3.2 words on average, eg, "Turn on the radio"} and b) human conversation\footnote{Longer phrases of 7.4 words on average with open vocabulary.}. 93 of the 195 speakers recorded a conversational test set, while all speakers recorded a home automation test set.
Speaker WER reported at each subsample size is an average across the 10 models personalized with the 10 random subsamples of the same size.

We chose different WER thresholds as success criteria for our two application scenarios: Home automation and conversational domains.
Voice assistants -- Apple's Siri, Amazon's Alexa, Microsoft's Cortana, and Google's Assistant being the most popular ones --  typically utilize Natural Language Understanding capabilities to infer user intent allowing them to function even without perfect transcriptions from the ASR system. We found that on Google Assistant, a WER of 15\% corresponds to a success rate of around 80\% for home automation commands. 
The conversational domain consists of less structured, longer and more complex phrases. Accordingly, we selected a higher WER target of 20\%, assuming that this renders transcriptions useful in a human conversation.

\section{Results}

\begin{table}[t]
  \centering
  \begin{tabular}{|l|l|l|l|l|l|}
  \hline
  &  & \multicolumn{4}{|c|}{target WER}\\
  domain & speakers &  5 & 10 & 15 & 20 \\
  \hline
  home automation & 195 & 59\% & 81\% & 90\% & 94\%\\
  conversational & 93 & 18\% & 59\% & 74\% & 82\%\\
  \hline
  \end{tabular}
  \caption{Success percentage (portion of speakers reaching target WER) when personalizing with all their data. }
  \label{tab:all_data_results}
\end{table}

We aim to understand the expected portion of users for which a personalized model trained on a specific amount of data will work in a satisfactory manner. We hence report \textbf{Success Percentage} -- the percentage of speakers to reach a domain specific target WER threshold.

Table~\ref{tab:all_data_results} shows the Success Percentage for four WER thresholds when using all the data available per speaker (as per Figure~\ref{fig:utt_count_boxplot}). The conversational domain is more challenging and percentages are significantly lower than for home automation. Yet, when using all data per speaker we reach a target WER of 15 for 90\% of the speakers on home automation, and a target WER of 20 for 82\% of the speakers on the conversational domain. These numbers serve as baselines.

\subsection{Percentage of speakers to reach a target WER}

\begin{figure}[t]
    \centering
    \includegraphics[width=0.47\textwidth]{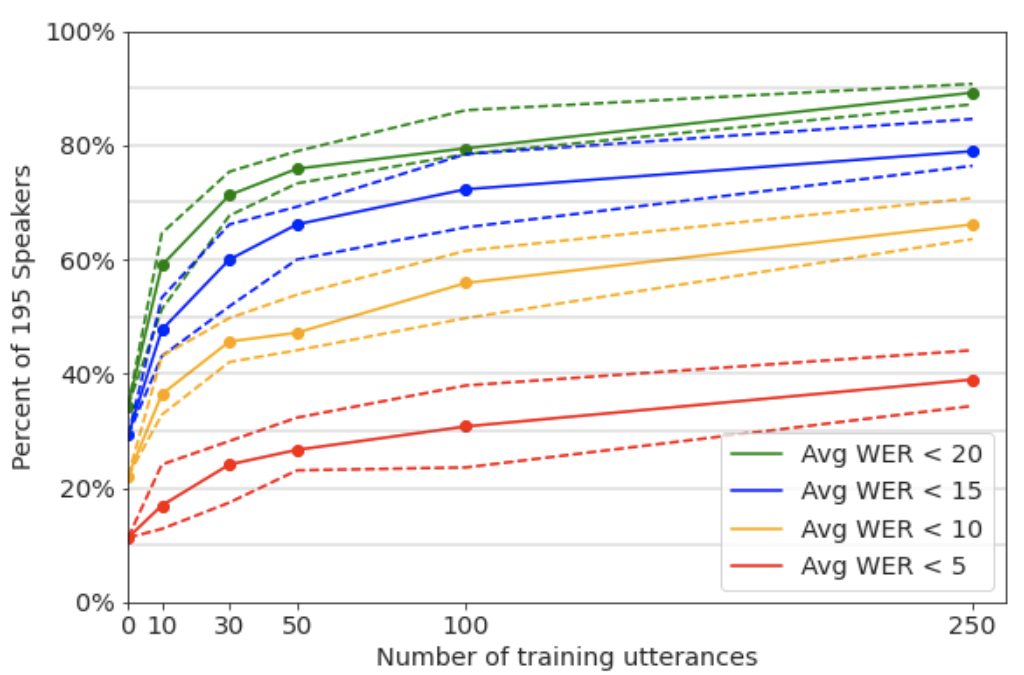}
    \caption{Success percentage on home automation domain, different target WERs (dashed lines: 95\% confidence interval).}
    \label{fig:assistant_by_target_wer}
\end{figure}

\begin{figure}[t]
    \centering
    \includegraphics[width=0.47\textwidth]{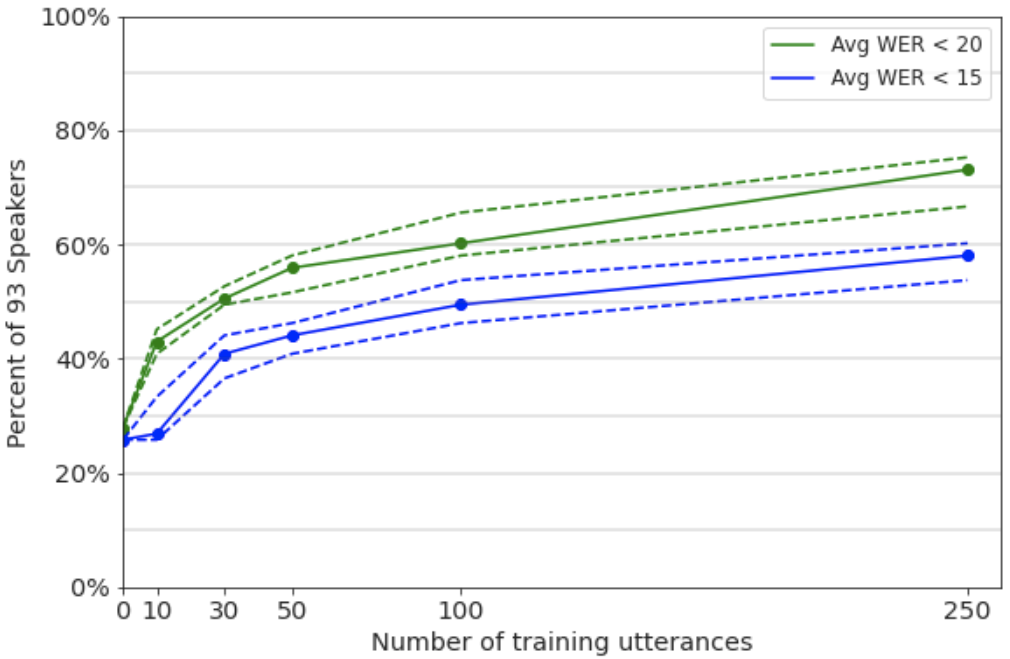}
    \caption{Success Percentage on conversational domain, different target WERs (dashed lines: 95\% confidence interval).}
    \label{fig:conversational_by_target_wer}
\end{figure}

Figures~\ref{fig:assistant_by_target_wer} and ~\ref{fig:conversational_by_target_wer} show Success Percentages for both domains for several different target WERs. In both cases, the choice of target WER significantly impacts the results. 
On the home automation domain for a target WER of 15, 250 utterances yield a Success Percentage of 79\%;  with as little as 50 utterances, the Success Percentage is 63\%. This marks a significant improvement compared to the unadapted, out-of-the-box model, which reaches this target WER for only 35\% of the speakers with speech impairment. 
On the conversational domain, Success Percentages are significantly lower and we restrict our analysis to two higher target WER scenarios. For a target WER of 20, the Success Percentage is about 73\% for speakers' models personalized on 250 utterances. 
This is a clear improvement compared to the base model which yields the same target WER only for 29\% of speakers.

Overall, this analysis shows that many speakers need only a small amount of data to achieve satisfactory model performance.
Many speakers will not need to record more than 250 utterances, with the biggest improvements coming from the first 50. Some users may still want to record more than 250 utterances in order to get sufficient improvement, albeit with marginal gains at a relatively high recording price (e.g. By recording an order of magnitude more data than 250 utterances, a median of 1788 utterances, the Success Percentage in the home automation domain increases by 11\% more speakers and 9\% more speakers in the conversational domain).

\subsection{Break-down by severity of speech impairment}

\begin{table}[h]
\centering
\begin{tabular}{|l|l|l|l|l|l|l|}
\hline
& num & \multicolumn{5}{|c|}{number of utterances} \\
  severity & spkrs & 0 & 10  &  50   & 250  & All   \\
  \hline
  Typical   & 15 & 100\% & 100\% & 100\% & 100\% & 100\% \\
  Mild & 69 & 45\% & 78\%  & 93\% & 96\% & 99\% \\
  Mod & 62 & 10\% & 32\% & 60\% & 81\% & 94\% \\
 Severe & 49 & 10\% & 8\% & 27\% & 47\% & 69\% \\
  \hline
  \end{tabular}
  \caption{Percentage of speakers reaching target WER of 15 on home automation domain, split by severity.}
  \label{tab:assistant_by_severity_table}
\end{table}

\begin{table}[h]
\centering
\begin{tabular}{|l|l|l|l|l|l|l|}
\hline
& num & \multicolumn{5}{|c|}{number of utterances} \\
  severity & spkrs & 0 & 10  &  50   & 250  & All   \\
  \hline
  Typical & 4 & 100\% & 100\% & 100\% & 100\% & 100\% \\
  Mild & 29 & 41\% & 62\%  & 86\% & 93\% & 97\% \\
  Mod & 37 & 16\% & 35\% & 46\% & 70\% & 86\% \\
 Severe & 23 & 17\% & 22\% & 26\% & 48\% & 52\% \\
  \hline
  \end{tabular}
  \caption{Percentage of speakers reaching target WER of 20\% on conversational domain, split by severity.}
  \label{tab:conversational_by_severity_table}
\end{table}

To further understand how much personalization data is really necessary for a given speaker, we here report Success Percentages
for the different levels of severity of speech impairment. For home automation, we chose a target WER of 15 (Table~\ref{tab:assistant_by_severity_table}), for conversational we chose a target WER of 20 (Table~\ref{tab:conversational_by_severity_table}). For all speakers which are rated as "typical" (i.e. no speech impairment), out-of-the-box ASR models work well with a maximum WER of 9 (home automation) and 12 (conversational). As severity increases, we observe that the maximum possible Success Percentage using all data decreases: 
only 52\% of speakers with a severe speech impairment on conversational and 69\% of those speakers on home automation. With even fewer training data  (especially $<$100 utterances), success percentages here sharply decline.

On the other hand, for speakers with a mild speech impairment, small amounts of training data  yield big Success Percentages, getting close to the rates achievable with all data: With only 50 utterances (i.e. about 3-4 minutes of recorded speech, Figure~\ref{tab:dataset}) we can expect 90\% of these speakers to yield the target WER of 15\%. This is not far off the 99\% Success Percentage of the "all data" scenario which clocks in with a median duration of 2.1 hours of speech recordings (1788 utterances) per speaker. These are large gains when contrasted with the 45\% success percentage of the out-of-the-box model (results on conversational domain analogous).

This analysis shows that the answer to the question of how much data is needed to achieve well working, personalized models for speakers with speech impairment clearly depends on the severity of impairment. Personalized models for speakers with moderate and severe impairment benefit significantly from more training data beyond the maximum subsampling size of 250 utterance of this study.
For mildly impaired speakers, however, the effort of recording more samples for personalization has only marginal benefit.

\subsection{Implications for spontaneous free-form speech}

\begin{figure}[tb]
    \centering

    \includegraphics[width=0.47\textwidth]{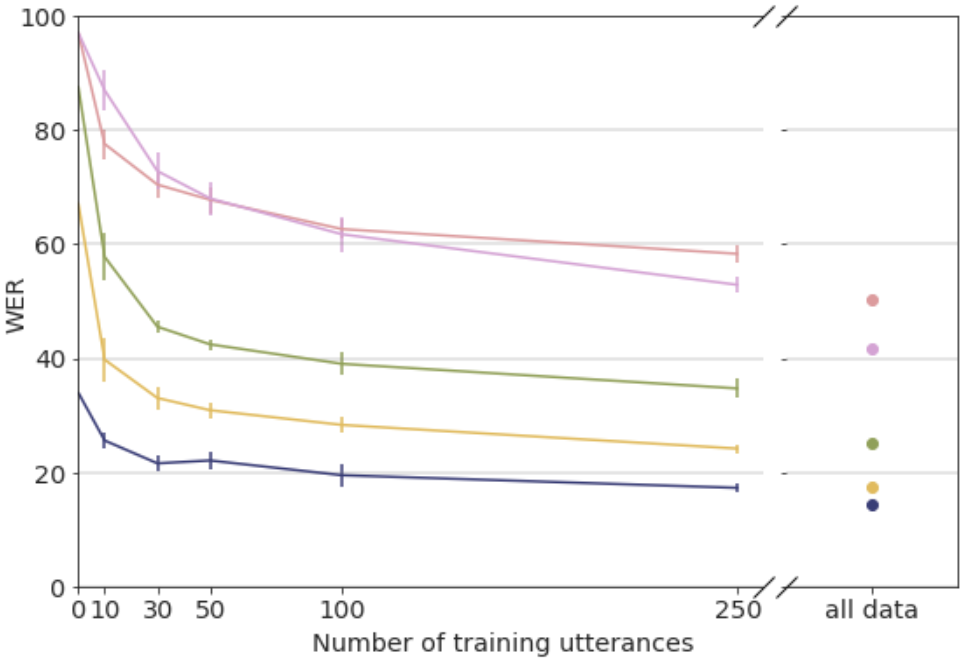}
    \caption{Average WER and 95\% confidence intervals on organic test set of 5 selected speakers. Last point is WER when training with all data of given speaker.}
    \label{fig:organic_selected_speakers}
\end{figure}

In the previous sections, both the training and the test data were taken from the Euphonia corpus where all recordings are based on prompted speech \cite{macdonald2021}. Our experiments clearly show that
useful personalized models can be obtained with relatively small amounts of training data
per speaker. However, an important, yet to answer question is whether such models also work in a scenario of spontaneous speech of an unknown domain. In other words, how well do models personalized on small amounts of data generalize to other domains, acoustic conditions, and scripted vs unscripted speech.

For 15 of the speakers included in our study, we recorded a so-called "organic test set". In direct interactions with these individuals, these speakers' spontaneous speech on a topic of their choice was recorded, broken into utterances and then transcribed by speech professionals. Speakers recorded an average of 140 test utterances. 
Figure~\ref{fig:organic_selected_speakers} shows the WER curves for five selected speakers.\footnote{Due to the small amount of speakers and the fact that the organic test sets of all speakers have different characteristics in domain and phrase complexity we refrain from showing Success Percentage for a target WER.}

Unsurprisingly, WERs on the organic test set are higher because spontaneous speech is more unlike the training set.

However, just as with the prompted home automation and conversational domains, we can see a knee in the WER curves within the first 100 utterances -- the biggest improvements are even gained within the first 50 utterances. Table \ref{tab:organic_rel_improvement} shows the relative WER improvement for different training set sizes averaged over all speakers. Note, that there is significant variance between the 15 speakers and we can see ongoing relevant gains beyond 250 utterances. Overall we observe the same trends as on prompted data, showing that training with small amounts of out-of-domain data does indeed generalize and leads to large model quality improvements on spontaneous speech.

\begin{table}[t]
  \centering
  \begin{tabular}{|r|r|}
  \hline
  utterances & average improvement (std dev) \\
  \hline
  avg = 3239 (std dev = 1717)      & 56\% (15\%) \\
  250        & 45\% (16\%) \\
  50        & 34\% (12\%) \\
  10        & 17\% (12\%) \\
  \hline
    \end{tabular}
  \caption{Average improvement over out-of-the-box model on organic test set across 15 speakers.}
  \label{tab:organic_rel_improvement}
\end{table}

\section{Conclusions}

Our study shows that small amounts of training data can indeed be sufficient to provide useful personalized ASR models for a large percentage of speakers with speech impairments, especially for use cases like home automation with limited vocabulary and less complex phrases. 
In our scenario, with as little as 17.8 minutes of recorded speech samples on average per speaker, we were able to produce personalized models with a pre-defined target WER for 96\% of speakers with mild, and 81\% of speakers with moderate speech impairment across a variety of etiologies and types of speech disorders.

This study also emphasizes how severity of speech impairment and amount of training data needed for useful models are correlated. For speakers with more severe speech impairment, we see larger gradual model improvements as the amount of training data is increased. 
We also show that even with small amounts of training data from narrow domains and prompted speech, personalized speech models show big improvements in recognition quality on out-of-domain phrases and spontaneous speech, suggesting good generalization properties of personalized models.

In future work, we aim to study how we can further reduce the amount of data needed for speech model personalization based on task-specific utterance selection.

\vfill\pagebreak
\bibliographystyle{IEEEbib}
\bibliography{refs}

\end{document}